\documentclass[12pt, draftclsnofoot, onecolumn]{IEEEtran}

\usepackage[mathcal]{euscript}
\usepackage{amssymb}
\usepackage{amssymb}
\usepackage{amsmath,bm,amsthm}
\usepackage{mnsymbol}
\usepackage{graphicx}
\usepackage{psfrag}
\usepackage{epstopdf}
\usepackage{epsf,epsfig}
\usepackage{enumitem}
\usepackage{subfigure}
\usepackage{cite,color}
\usepackage{mathrsfs}
\usepackage{multirow}
\usepackage{multicol}
\usepackage{stfloats}
\usepackage{makecell}
\usepackage{tabularx}
\def\BibTeX{{\rm B\kern-.05em{\sc i\kern-.025em b}\kern-.08em T\kern-.1667em\lower.7ex\hbox{E}\kern-.125emX}}
\allowdisplaybreaks
\renewcommand\IEEEkeywordsname{Index Terms}
\begin{document}

\title{Permutation Polynomial Interleaved Zadoff-Chu Sequences}

\author{Fredrik Berggren, \IEEEmembership{Senior Member, IEEE} and Branislav M. Popovi\'{c}
\thanks{The authors are with Huawei Technologies Sweden AB, Box 54, SE-164 94 Kista, Sweden.}}

\maketitle
%\vspace{-20mm}
\begin{abstract}
Constant amplitude zero autocorrelation (CAZAC) sequences have modulus one and ideal periodic autocorrelation function. Such sequences are used in cellular radio communications systems, e.g., for reference signals, synchronization signals and random access preambles. We propose a new family CAZAC sequences, which is constructed by interleaving a Zadoff-Chu sequence by a quadratic permutation polynomial (QPP), or by a permutation polynomial whose inverse is a QPP. It is demonstrated that a set of orthogonal interleaved Zadoff-Chu sequences can be constructed by proper choice of QPPs. 

\end{abstract}
%\vspace{-5mm}
%\vspace{-5mm}
\section{Introduction}
\IEEEPARstart{C}{onstant} amplitude zero-autocorrelation (CAZAC) sequences are frequently used in waveforms for radar and for communications systems \cite{Ye}, as they provide constant envelope and have ideal periodic autocorrelation function. CAZAC sequence constructions include, e.g., Zadoff-Chu sequences \cite{Chu}, Frank sequences \cite{Frank}, generalized chirp-like (GCL) sequences \cite{Popovic}, Björck sequences \cite{Bjorck}, extended Björck sequences \cite{Popovicx}, cubic polynomial phase sequences \cite{Song}, biphase sequences \cite{Bomer} and modulatable CAZAC sequences \cite{Popovic3}. Moreover, exhaustive search by computers have been performed to find CAZAC sequences. If the sequence length $N$ is non-prime and divisible by a perfect square, then there are infinitely many CAZAC sequences. If $N$ is non-prime and not divisible by a perfect square, it is unknown how many CAZAC sequences there are \cite{Magsino}. 

It has been shown that the CAZAC property is invariant under certain basic mathematical operations \cite{Benedetto1}. Herein, we are considering permutations and we are particularly interested in the properties of a CAZAC sequence being interleaved by a permutation polynomial. Interleavers based on permutation polynomials have been used for turbo codes \cite{Takeshita}\cite{Rosnes} and for improving performance of single-carrier waveforms over fading radio channels \cite{Berggren}. 
However, it appears that prior work has not provided insight on how to determine when, or whether, the CAZAC property is preserved for an arbitrary permutation, or for which CAZAC sequences it could be preserved. Zadoff-Chu sequences are frequently used in practice, e.g., as random access preambles, synchronization signals and reference signals in 4G and 5G \cite{38211}, and generating large sets of sequences with good correlation properties is a well-known problem. By allocating signals from different sets to cells and users, the inter-cell and intra-cell interference can be suppressed. Therefore, in this work, we will consider permutation polynomial interleaved Zadoff-Chu sequences, as a means to produce more CAZAC sequences. 
The necessary definitions are given in Sec. II and the main results are stated in Sec. III. Examples of sets of sequences are given in Sec. IV while Sec. V concludes the paper.

\section{Definitions}
Define the periodic autocorrelation function (PACF) of sequence $x[k], k=0,1,\ldots, N-1$ for delays $d=0,1,\ldots, N-1$ 
\begin{align}
\theta(d) &=\sum_{k=0}^{N-1}x[k]x^*[k+d\, (\mathrm{mod}\, N)] 
\label{eq:PACF}
\end{align}
where $(\mathrm{mod}\, N)$ is the modulo-$N$ operator and $(\cdot)^*$ denotes complex conjugate.
A CAZAC sequence has constant amplitude, $|x[k]|=1, \forall k$, and ideal PACF, i.e., $\theta(d)=N\delta[d]$, where the Kronecker delta function is $\delta[0]=1$ and $\delta[k]=0$ for $k\neq 0$. If $x[k]$ is a CAZAC sequence, then $y[k]$ is a CAZAC sequence under the following basic mathematical operations \cite{Benedetto1} \cite{Benedetto}:
\begin{enumerate}[label=\roman*.]
\item (Rotation) $y[k]=cx[k]$, for some $|c|=1$
\item (Translation) $y[k]=x[k+f_0 (\mathrm{mod}\, N)]$, where $f_0$ is an integer
\item (Decimation) $y[k]=x[f_1k (\mathrm{mod}\, N)]$, where $f_1$ is an integer being relatively prime to $N$
\item (Linear frequency modulation) $y[k]=W_N^{kn}x[k]$, where $n$ is an integer and $W_N=e^{-\mathrm{j}\frac{2\pi}{N}}$ with $\mathrm{j}=\sqrt{-1}$
\item (Conjugation) $y[k]=x^*[k]$
\end{enumerate}

A Zadoff-Chu sequence can be defined by \cite{Popovic}
\begin{align}
x[k]=W_N^{u(k+(N (\mathrm{mod}\, 2))+2q)k/2}
\label{eq:ZC}
\end{align}
where $u$ is an integer, aka root index, which is relatively prime to $N$. It could be noted that the term $q$ in (\ref{eq:ZC}) amounts to a linear frequency modulation. We will consider the interleaved Zadoff-Chu sequence 
\begin{equation}
y[k]=x[\pi[k]]
\label{eq:zcinter}
\end{equation}
where 
\begin{equation}
\pi[k]=f_v k^v+f_{v-1} k^{v-1}+\ldots+f_0 \, (\mathrm{mod} N)
\label{eq:pipol}
\end{equation}
is a $v$th degree permutation polynomial, i.e., it produces a permutation of the values in the set $\{0,1,\ldots,N-1\}$ and the coefficients $f_i\in\{0,1,\ldots,N-1\}$. There exists at least one inverse permutation polynomial, $\pi^{-1}[k]$, to each permutation polynomial, such that $\pi^{-1} [\pi[k]]=k$, cf. \cite{Lahtonen},\cite{Nieminen},\cite{Ryu}. The degree of $\pi^{-1}[k]$ may not be the same as for $\pi[k]$. For $v=1$, i.e., a linear permutation polynomial (LPP), it is required that $\mathrm{gcd}(f_1,N)=1$, where $\mathrm{gcd}(a,b)$ denotes the greatest common divisor of $a$ and $b$. For $v=2$ and $v=3$, i.e., quadratic permutation polynomial (QPP) and cubic permutation polynomial (CPP), conditions on the coefficients can be found in \cite{Takeshita}\cite{Chen}. In general, any $N$ can be represented as $N=p_0^{n_0} p_1^{n_1} \ldots  p_r^{n_r}$, where $p_i, i=0,1,\ldots, r$ are distinct prime numbers and the multiplicities $n_i\,(n_i>0)$ are integers. From \cite{Trifina}, the conditions on the QPP coefficients are:
\begin{align}
\mathrm{If}\, p_i=2,n_i=1& \,\mathrm{then} \, f_1+f_2 \nequiv 0 \,(\mathrm{mod} \,2).\label{eq:qppc1}\\	
\mathrm{If}\, p_i=2,n_i>1& \,\mathrm{then}\, f_1\nequiv 0 \,(\mathrm{mod} \,2)\, \mathrm{and} \, f_2\equiv 0  \,(\mathrm{mod} \,2).\label{eq:qppc2}\\
\mathrm{If}\, p_i>2,n_i\ge 1& \, \mathrm{then}\, f_1\nequiv 0 \,(\mathrm{mod} \,p_i)\, \mathrm{and}\, f_2\equiv 0  \,(\mathrm{mod} \,p_i).\label{eq:qppc3}
\end{align}
These conditions may not be feasible for certain $N$, e.g., for prime numbers. Irreducible QPPs produce permutations which cannot be obtained from an LPP. Irreducible QPPs do not exist for all feasible $N$ \cite{Zhao} and it has been shown that a QPP is irreducible if and only if $\mathrm{gcd}(N,2f_2)<N$ \cite{Takeshita}. According to \cite{Trifina}, when the prime factorization of $N$ is such that $p_0=2$, $n_0=0, \mathrm{or} \, 1, \mathrm{or} \, 2$ and $n_i=1, i=1,2,...,r$, then there are no irreducible QPPs. Two different permutation polynomials can generate the same permutation and the number of irreducible QPPs producing unique permutations depends on $N$, which can be computed by formulas given in \cite{Trifina},\cite{Zhao}. 

\section{Zadoff-Chu Sequences Interleaved by Permutation Polynomials}
\subsection{CAZAC Property}
For an LPP, it is straightforward to show that (\ref{eq:zcinter}) will be a CAZAC sequence, since $f_0$ and $f_1$ correspond to the translation and decimation basic mathematical operations, respectively. Thus, we will focus on irreducible permutation polynomials with a degree $v \ge 2$. 

{\it Theorem 1:} If $x[k], k=0,1,\ldots, N-1$ is a Zadoff-Chu sequence and $\pi[k]$ is a QPP, then $x[\pi[k]]$ is a CAZAC sequence.

\emph{Proof:} Since $f_0$  is a translation, which is a mathematical operation that preserves the CAZAC property, we can set $f_0=0$ in the proof. Expanding (\ref{eq:zcinter}) with a QPP with $f_0=0$ produces 
\begin{align}
y[k]&=W_N^{u(f_2^2k^4+2f_1f_2k^3+(f_1^2+f_2(N\,(\mathrm{mod} \,2)+2q))k^2)/2)/2} \notag \\
&\times W_N^{u(f_1k(N\,(\mathrm{mod} \,2)+2q))/2}\label{eq:zcp}
\end{align}
so we can rewrite (\ref{eq:zcp}) as 
\begin{align}
y[k]&=W_N^{-u(q+(N\,(\mathrm{mod} \,2))q/2}\notag \\
&\times W_N^{u(f_2k^2+f_1k+q+(N\,(\mathrm{mod} \,2))(f_2k^2+f_1k+q)/2}
\end{align}
which implies that $q$ introduces a rotation and a translation, which are mathematical operations that preserve the CAZAC property. Hence, we can set $q=0$ in the proof. Inserting (\ref{eq:zcp}) with $q=0$ in (\ref{eq:PACF}) gives the PACF
\begin{align}
\theta(d)&=C_0 \sum_{k=0}^{N-1}W_N^{-u(g_3k^3+g_2k^2+g_1k)} \label{eq:pacf}\\
C_0&=W_N^{-u\left(f_2^2d^4+2f_1f_2d^3+f_1^2d^2 \right)/2}W_N^{-u\left(f_2d^2+f_1d\right)/2    (N\,(\mathrm{mod} \,2)) }\\
g_3&=2f_2^2d \label{eq:g3}\\
g_2&=3f_2d(f_2d+f_1) \label{eq:g2}\\
g_1&=d(2f_2d+f_1)(f_2d+f_1)+df_2(N\, (\mathrm{mod}\, 2)). \label{eq:g1a}
\end{align}
We will show that $\theta(d)=N\delta[d]$. First, we can rewrite (\ref{eq:pacf}) 
\begin{equation}
\theta(d)=C_0 \sum_{k=0}^{N-1}W_N^{-u(g_3(k+t)^3+g_2(k+t)^2+g_1(k+t))} 
\label{eq:pacf2}
\end{equation}
for any integer $t$, because the sum is independent of the order of its terms and since $W_N^{-ug_sk^s}=W_N^{-ug_s(k+N)^s}$ for $s=1,2,3$. Expanding (\ref{eq:pacf2}) gives
\begin{align}
\theta(d)&=C_0 \sum_{k=0}^{N-1}W_N^{-u(g_3k^3+g_2k^2+g_1k)}W_N^{-u(g_3t^3+g_2t^2+g_1t)} \notag \\
&\times W_N^{-ug_3(3kt^2+3k^2t)} W_N^{-ug_22kt}. \label{eq:pacfe}
\end{align}
We shall show that when $d\neq0$, there exists an integer $t=t_c \,(1\le t_c \le N-1)$ which causes that 
\begin{equation}
C_1=W_N^{-u(g_3t_c^3+g_2t_c^2+g_1t_c)}W_N^{-ug_3(3kt_c^2+3k^2t_c)} W_N^{-ug_22kt_c}
\label{eq:sumt}
\end{equation}
becomes  a constant $C_1\neq 1$ for $\forall k, 0\le k \le N-1$. Thus, according to (\ref{eq:pacf}) and (\ref{eq:pacfe}) 
\begin{equation}
\theta(d)=C_1\theta(d)
\end{equation}
which implies that the solution is $\theta(d)=0$ for $d\neq 0$. 
To prove that $C_1\neq 1$ when $d\neq 0$ assume now that 
\begin{equation}
t_c=
\begin{cases}
N/\mathrm{gcd}(2g_2,N), & N\equiv 0 \, (\mathrm{mod}\, 2), N\nequiv 0 \, (\mathrm{mod}\, 4),\\
& f_2 \equiv 1\, (\mathrm{mod}\, 2), d\equiv 1 \, (\mathrm{mod}\, 2),\\
N/\mathrm{gcd}(g_2,N), & \mathrm{otherwise.}
\end{cases}
\end{equation}
According to Lemma 2 and Lemma 3 in Appendix A, $g_3t_c \equiv 0\, (\mathrm{mod}\, N)$ and either $g_1t_c \nequiv 0\, (\mathrm{mod}\, N)$ or
$g_2t_c^2+g_1t_c \nequiv 0\, (\mathrm{mod}\, N)$, so we have 
\begin{equation}
g_3t_c^3+g_2t_c^2+g_1t_c \nequiv 0 \, (\mathrm{mod}\, N)
\label{eq:cond1a}
\end{equation}
which means that 
\begin{equation}
W_N^{-u(g_3t_c^3+g_2t_c^2+g_1t_c)}\neq 1.
\label{eq:cond1}
\end{equation}
By defining 
$z[k]=W_N^{-ug_3(3kt_c^2+3k^2t_c)} W_N^{-ug_22kt_c}$, we have
\begin{align}
\frac{z[k+1]}{z[k]}&=W_N^{-u(g_3(3t_c^2+3t_c+6kt_c)+g_22t_c)}\\
&=W_N^{-ug_36kt_c}W_N^{-u(g_3(3t_c^2+3t_c)+g_22t_c)}. \label{eq:sq}
\end{align}
As we established above, $g_3t_c \equiv 0\, (\mathrm{mod}\, N)$, so 
\begin{equation}
g_36kt_c\equiv 0 \, (\mathrm{mod}\, N)
\label{eq:cond2a}
\end{equation}
which implies that $W_N^{-ug_36kt_c}=1$.
We further have that 
\begin{align}
g_3(3t_c^2+3t_c)+g_22t_c &\equiv 3t_cg_3(t_c+1)+g_22t_c  \, (\mathrm{mod}\, N)\\
&\equiv g_22t_c \, (\mathrm{mod}\, N)\\
& \equiv 0 \, (\mathrm{mod}\, N). 
\end{align}
as either $g_2t_c \equiv 0\, (\mathrm{mod}\, N)$ or $2g_2t_c \equiv 0\, (\mathrm{mod}\, N)$ according to Lemma 2 and Lemma 3.
Hence, we have $W_N^{-u(g_3(3t_c^2+3t_c)+g_22t_c)}=1$ and therefore $z[k+1]=z[k]$. Since $z[0]=1$, we have $z[k]=1$, so 
from (\ref{eq:sumt}) and (\ref{eq:cond1}) it follows that $C_1$ is a constant where $C_1\neq1$.$\hfill \blacksquare$

{\it Theorem 2:} If $x[k], k=0,1,\ldots, N-1$ is a Zadoff-Chu sequence and $\pi[k]$ is a permutation polynomial and its inverse is a QPP, $\pi^{-1}[k]=h_2k^2+h_1k+h_0 \,(\mathrm{mod} \,N)$,  or an LPP, $\pi^{-1}[k]=h_1k+h_0 \,(\mathrm{mod} \,N)$, then $x[\pi[k]]$ is a CAZAC sequence.

\emph{Proof:} Since $x[\pi[k]]$ is a CA sequence, we only need to show its ZAC property. Define the discrete Fourier transform (DFT) sequence
\begin{equation}
X[k]=\frac{1}{\sqrt{N}} \sum_{m=0}^{N-1}x[\pi[m]]W_N^{mk} 
\label{eq:Xa}
\end{equation}
and consider Proposition 2.1 in \cite{Benedetto}, which implies that $x[\pi[m]]$ is a ZAC sequence if and only if $X[k]$ is a CA sequence, which is shown as follows, 
\begin{align}
X[k]X^*[k]&=\frac{1}{\sqrt{N}} \sum_{m=0}^{N-1}x[\pi[m]] W_N^{mk} \notag \\
& \times \frac{1}{\sqrt{N} }\sum_{p=0}^{N-1}x^* [\pi[p]] W_N^{-pk}  \notag \\
&\stackrel{\mathrm{(a)}}{=}\frac{1}{N} \sum_{m=0}^{N-1}x[m] W_N^{\pi^{-1} [m]k}  \sum_{p=0}^{N-1}x^* [p] W_N^{-\pi^{-1} [p]k} \notag\\
&\stackrel{\mathrm{(b)}}{=}\frac{1}{N} \sum_{m=0}^{N-1}x[m] W_N^{\pi^{-1} [m]k} \notag \\
&\times \sum_{v=0}^{N-1}x^* [v+m \, (\mathrm{mod}\,N)] W_N^{-\pi^{-1} [v+m]k} \notag \\
&\stackrel{\mathrm{(c)}}{=}\frac{1}{N} \sum_{m=0}^{N-1}x[m] W_N^{\pi^{-1} [m]k} \notag \\
&\times \sum_{v=0}^{N-1}x^* [v+m] W_N^{-( \pi^{-1} [v]+\pi^{-1}[m]+2h_2vm-h_0)k} \notag\\
&\stackrel{\mathrm{(d)}}{=}\frac{W_N^{h_0k}}{N} \sum_{v=0}^{N-1}W_N^{-\pi^{-1} [v]k} \notag \\
&\times \sum_{m=0}^{N-1}x[m]x^* [v+m] W_N^{- 2h_2 vmk}  \notag\\
&\stackrel{\mathrm{(e)}}{=}\frac{W_N^{h_0k}}{N} \sum_{v=0}^{N-1}W_N^{-\pi^{-1} [v]k} \notag \\
& \times W_N^{-uv(v/2+ (N\, (\mathrm{mod} 2))/2+q)}  \sum_{m=0}^{N-1}W_N^{- vm(u+2kh_2 )} \notag \\
&\stackrel{\mathrm{(f)}}{=}W_N^{h_0k}W_N^{- \pi^{-1} [0]k} \notag \\
&\stackrel{\mathrm{(g)}}{=}1. \notag
\end{align}
Step (a) follows from $\pi^{-1} [\pi[m]]=m$. Step (b) is obtained by introducing the change of variable $p=v+m$, which just reorders the terms of the sum over $p$. Step (c)  follows from the definition of $\pi^{-1}[k]$. Step (d) is obtained by changing the order of summations. Step (e) follows from the definition of $x[m]$. Step (f) follows from the fact that $u+2kh_2 \nequiv 0 \, (\mathrm{mod} N)$ according to Lemma 4 in Appendix B, which makes the inner sum in step (e) equal to zero for any non-zero $v$.
Step (g) follows from $\pi^{-1} [0]=h_0$. The above steps can be applied also when $\pi^{-1}[k]=h_1k+h_0\, (\mathrm{mod} N)$, i.e., for an LPP.$\hfill \blacksquare$

It should be remarked that the requirement that $\pi^{-1}[k]$ is a QPP or an LPP is a sufficient but not a necessary condition for an interleaved Zadoff-Chu sequence to retain the CAZAC property. 

{\it Example 1:} It can be verified by computing (\ref{eq:PACF}) with (\ref{eq:zcinter}), that using the irreducible CPP $\pi[k]=8k^3+2k^2+k \,(\mathrm{mod} \,32)$ results in a CAZAC sequence. This CPP has two QPP inverses: $\pi^{-1}[k]=6k^2+k \,(\mathrm{mod} \,32)$ and  $\pi^{-1}[k]=22k^2+17k \,(\mathrm{mod} \,32)$. Moreover, using the irreducible CPP $\pi[k]=2k^3+k \,(\mathrm{mod} \,32)$ also results in a CAZAC sequence. However, it does not have a QPP inverse, but 4 CPP inverses: $\pi^{-1}[k]=10k^3+17k \,(\mathrm{mod} \,32)$, $\pi^{-1}[k]=10k^3+16k^2+k \,(\mathrm{mod} \,32)$, $\pi^{-1}[k]=26k^3+k \,(\mathrm{mod} \,32)$, $\pi^{-1}[k]=26k^3+16k^2+17k \,(\mathrm{mod} \,32)$.

Based on Theorem 2, we can formulate the following construction of CAZAC sequences.

{\it Corollary 1:} If $x[k], k=0,1,\ldots, N-1$ is a Zadoff-Chu sequence and $\pi[k]$ is a QPP with an inverse permutation polynomial $\pi^{-1}[k]$, then $x[\pi^{-1}[k]]$ is a CAZAC sequence.

We have exhaustively evaluated (\ref{eq:PACF}) with (\ref{eq:zcinter}) to determine whether the CAZAC property is preserved for CPPs in general. We found that for certain $N$ all CPPs work, but for some $N$ no CPP works, and for other $N$ a subset of the CPPs works. 
Table \ref{table:perms} shows how many CPPs there are in total, how many unique permutations these CPPs generate, how many of these unique CPP permutations preserve the CAZAC property, and how many permutations out of total $N!$ permutations preserve the CAZAC property. 

\begin{table*}
\caption{Total number of CPPs, number of unique CPP permutations, number of unique CPP permutations which preserve the CAZAC property and number of permutations out of $N!$ which preserve the CAZAC property.}
%\vspace{-5mm}
\begin{center}
\label{table:perms}
\begin{tabular}{|c|c|c|c|c|}
\hline
\hline
$N$ & Total \#CPPs & \#CPP permutations& \#CPP CAZAC permutations & \#CAZAC permutations \\
\hline
3 & 12 &6 & 6 &6 \\
\hline
4 & 16 & 8&8 &8 \\
\hline
5 & 100&100 & 20 &40 \\
\hline
6 & 120&12&12&24 \\
\hline
7 & 0 &0 & 0 &168 \\
\hline
8  &384 & 128& 128&256\\
\hline
9 &810 &324& 324 &2592\\
\hline
10 &880&240&80&320\\
\hline
11 &1210&1210&0 &1760\\
\hline
12 & 480& 48& 48 &6912\\
\hline
\hline
\end{tabular}
\end{center}
%\vspace{-10mm}
\end{table*}

Furthermore, it should be noted that there exist QPPs with inverse permutation polynomials of degree larger than of a CPP. For example, QPPs with quartic ($v=4$) inverse permutation polynomial were given in \cite{Rosnes}. These could also be used according to Corollary 1.

\subsection{Uniqueness of the Interleaved Zadoff-Chu Sequences}
The uniqueness of a permutation polynomial interleaved Zadoff-Chu sequence can in general not be guaranteed. The first reason for this is the term $k^2$ in its exponent.
 
{\it Example 2:} If $N=8$, the QPPs $\pi_0[k]=2k^2+k\,(\mathrm{mod} \,8)$ and $\pi_1[k]=2k^2+3k\,(\mathrm{mod} \,8)$ generate two distinct permutations, $\pi_0[k]\nequiv \pi_1[k], k\neq 0,4$. However, it can be verified that $\pi_0^2[k]\equiv \pi_1^2[k]\,(\mathrm{mod} \,8), \forall k$, which implies that $x[\pi_0[k]]=x[\pi_1[k]]$. 

A second reason for this is the inherent central symmetry of the Zadoff-Chu sequences when $q=0$, such that $x[k]=x[N-k]$ when $N$ is even and $x[k]=x[N-1-k]$ when $N$ is odd. Hence, even if $\pi_0^2[k]\nequiv \pi_1^2[k]$, it is still possible that $x[\pi_0[k]]=x[\pi_1[k]]$. The central symmetry also implies that there could potentially be very many permutations, which are not obtained through QPPs or CPPs, that preserve the CAZAC property.

Furthermore, it is possible that a QPP interleaved Zadoff-Chu sequence becomes equivalent to a non-interleaved but manipulated Zadoff-Chu sequence, e.g., obtained by the basic mathematical operations listed in Sec. II. That is, there may exist a root index $u_2:\mathrm{gcd}(u_2,N)=1$, a translation $d\in \{0,1,\ldots,N-1\}$, a rotation $a\in\{0,1,\ldots,2N-1\}$, a linear frequency  modulation sequence $W_N^{vk}, v=0,1,\ldots,N-1$ and a complex conjugate operation $s\in\{-1,1\}$ such that 
\begin{align}
W_N^{u_1(f_2k^2+f_1k+(N\,(\mathrm{mod}\,2)))(f_2k^2+f_1k)/2}&=\notag \\
W_N^{vk}W_N^{su_2(k+d+(N\,(\mathrm{mod}\,2)))(k+d)/2+a/2}, k=0,1,\ldots,N-1. \label{eq:eqcond}
\end{align} 

{\it Example 3:} For $N=8$, it can be verified by evaluating (\ref{eq:eqcond}) that there are 12 QPPs and 8 of them generate interleaved Zadoff-Chu sequences which cannot be obtained from the basic operations of (\ref{eq:eqcond}): $f_2\in\{2,6\}$ and $f_1\in\{1,3,5,7\}$.

We have exhaustively evaluated (\ref{eq:eqcond}) with all combinations of $u_1,u_2,d,a,v,s$ for $N\le 128$ to determine whether the QPP interleaved sequences are unique. The results are contained in Fig. 1 and Fig. 2. In Fig. 1, we show the fraction of QPPs for which the interleaved sequence is unique compared to the total number of QPPs. For many $N$, a substantial amount of QPPs give unique sequences. For some $N$ all the QPPs give unique sequences. We can identify these values as $N\in\{25,49,121,125\}$ and we additionally could evaluate and verify that $N\in\{169,343\}$ also only give unique sequences. Thus, we conjecture that only unique sequences are obtained when $N=p^n$ where $p\, (p>3)$ is a prime number and $n\, (n>1)$ is an integer. There is also a set of values $N$ for which none of the interleaved sequences are unique, i.e., $N\in\{9,18,36,45,63,90,99,117,126\}$.

Fig. 2 shows the maximum number of unique sequences $x[\pi[k]]$ which can be obtained from different QPPs (with $f_0=0$) and a single root index $u$, as well as the maximum number of sequences which can be obtained from a single non-interleaved Zadoff-Chu sequence by using all feasible root indices. This demonstrates that for many $N$, QPP interleaving results in more unique sequences.  

\begin{figure}
\begin{centering}
\includegraphics[width=\linewidth]{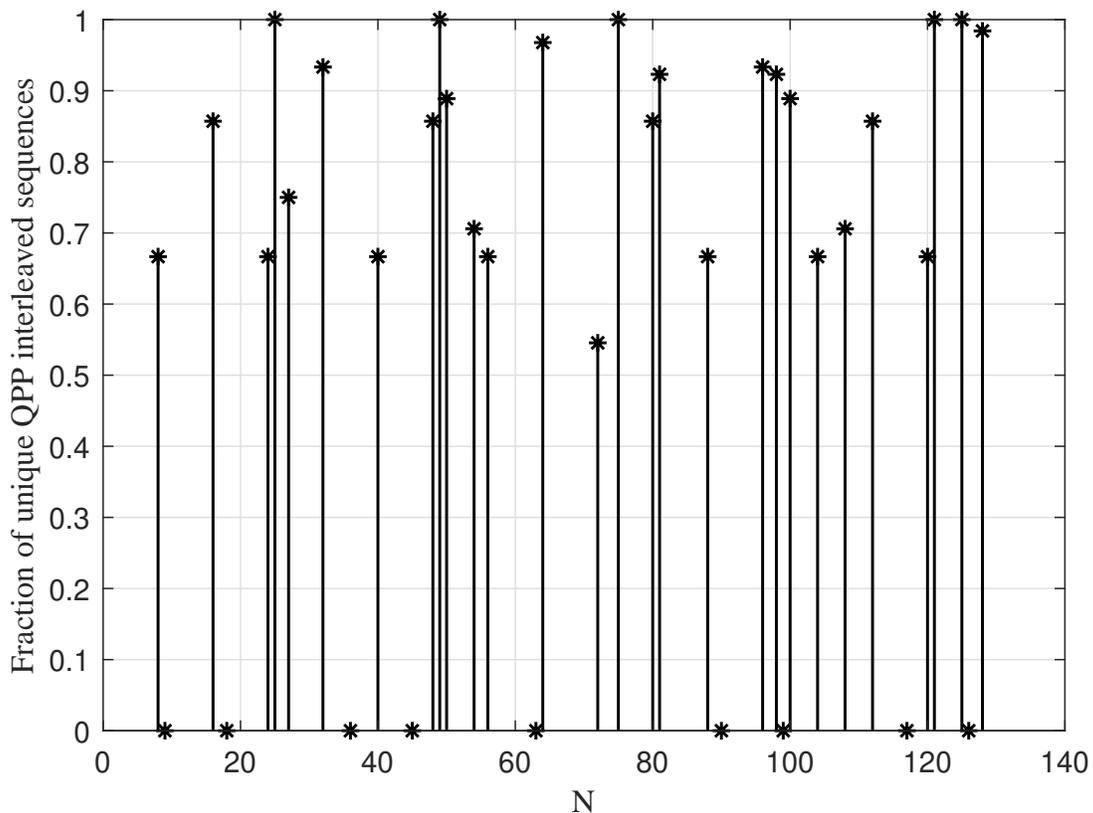}
\caption{Fraction of QPPs which produce an interleaved Zadoff-Chu sequence which cannot be obtained by basic mathematical operations on a non-interleaved Zadoff-Chu sequence.}
\end{centering}
\label{fig:set}
\end{figure}

\begin{figure}
\begin{centering}
\includegraphics[width=\linewidth]{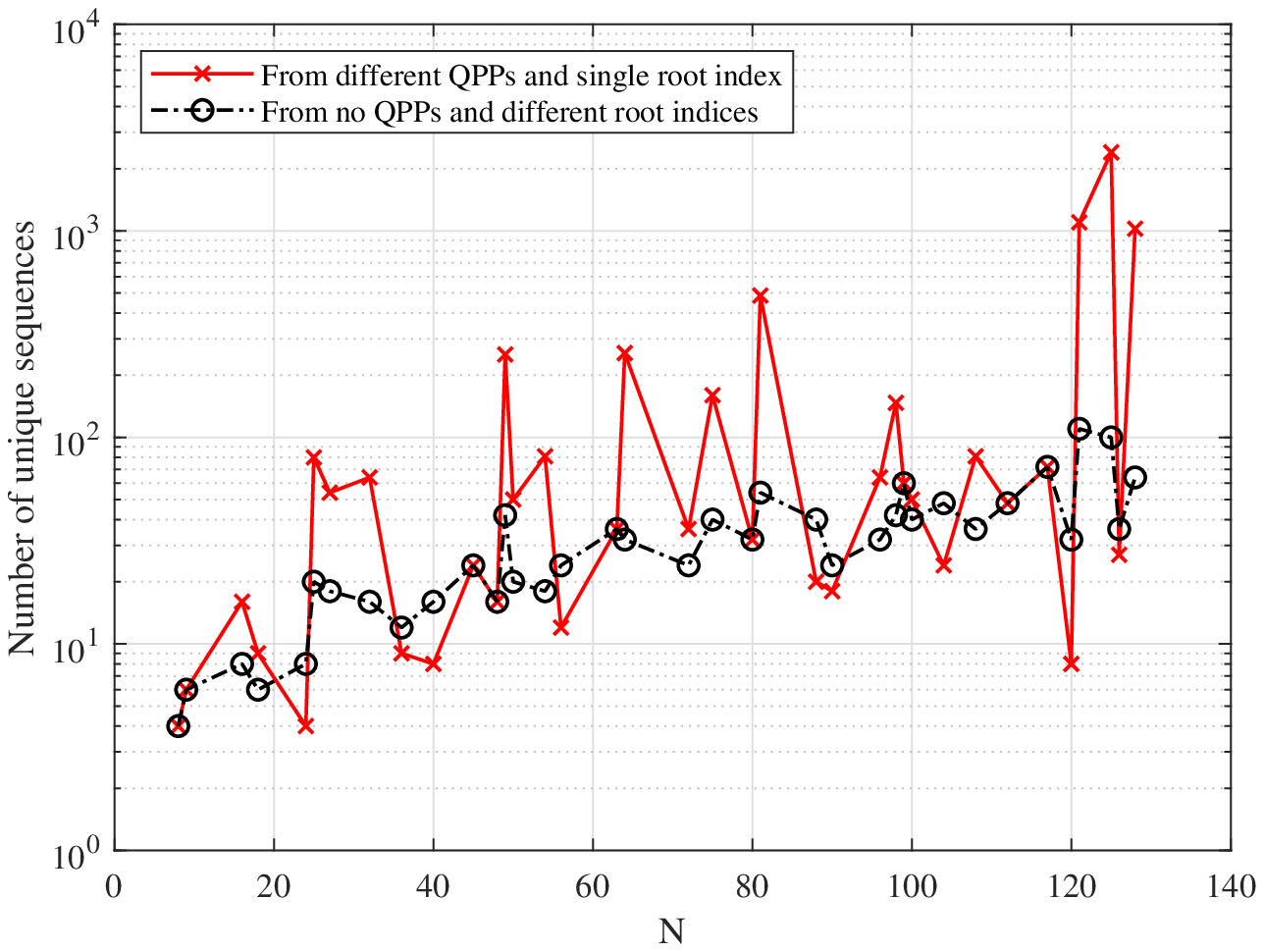}
\caption{Maximum number of unique sequences constructed by using different QPPs or by not using any QPPs but using different root indices $u$.}
\end{centering}
\label{fig:set}
\end{figure}

\subsection{Relation to GCL Sequences}
A further question is the relation between the interleaved Zadoff-Chu sequences and GCL sequences, i.e., modulated Zadoff-Chu sequences. We can equivalently write (\ref{eq:zcinter}) as a GCL sequence 
\begin{equation}
y[k]=w[k]x[k]
\label{eq:gcl}
\end{equation}
where the modulation sequence is 
\begin{equation}
w[k]=x[\pi[k]]x^*[k]. \label{eq:mods}
\end{equation}
For $q=0$, $w[k]$ is given by (\ref{eq:wmod}).
Moreover, from (\ref{eq:wmod}), we can decompose $w[k+\delta]$ as (\ref{eq:wperiodt}) for any integer $\delta$.
If $w[k]$ has period $\delta=T$, it is required that $w[k+T]=w[k]$.
\begin{figure*}[tbp]
\hrulefill
\begin{equation}
w[k]=
W_N^{u(f_2^2k^4+2f_1f_2k^3+(2f_2f_0+f_1^2-1)k^2+(2f_1f_0+(f_1-1)(N\,(\mathrm{mod}\,2))k+f_0(f_0+(N\,(\mathrm{mod}\,2))))/2}
\label{eq:wmod}
\end{equation}
%\hrulefill
\begin{equation}
w[k+\delta]=
w[k]w[\delta]		W_N^{u(2\delta f_2k^3+3f_2\delta(f_2\delta+f_1)k^2+\delta(2f_2\delta^2+3f_1f_2\delta+f_1^2-1+2f_2f_0)k-f_0(f_0+(N\,(\mathrm{mod}\,2))))}											
\label{eq:wperiodt}
\end{equation}
\hrulefill
\end{figure*} 
In\cite{Popovic}, it has been shown that if $N=sm^2$, where $s$ and $m$ are integers, then (\ref{eq:gcl}) is a CAZAC sequence for any unimodular sequence $w[k\,(\mathrm{mod} \,m)]$ with period $T=m$. The modulation sequences (\ref{eq:mods}) based on QPP can have period $T \neq m$ and are thus not necessarily equal to the construction in \cite{Popovic}. 

{\it Example 4:} Consider $N=2^{n_0}$ with $f_2=2^v, 0<v<n_0$, $f_1=1$ and $f_0=0$ as the coefficients of a QPP. If $n_0=5$, inserting $\delta=N/f_2$ in (\ref{eq:wperiodt}) results in $w[k+\delta]=w[k]$, i.e., the period of $w[k]$ is $T=2^{n_0-v}$, and thus $T \in \{2,4,8,16\}$. On the other hand, the possible values of $m$ are $m \in \{2,4\}$ since $N=8\cdot 2^2=2\cdot 4^2$.

\section{Set of Orthogonal Interleaved Zadoff-Chu Sequences}
From the decomposition (\ref{eq:gcl}) a set of orthogonal interleaved Zadoff-Chu sequences $\{y_i[k], 0\leq i \leq I-1 \}$ obtained by different QPPs can be defined as
\begin{equation}
     y_i[k]=w_i [k] x[k]. \label{eq:setgcl}  
\end{equation}
From (\ref{eq:setgcl}) it follows that the periodic cross-correlation function (PCCF) of two interleaved Zadoff-Chu sequences can be expressed as [23, Eq. (16)]
\begin{align}
\theta_{ij}(d)&=\sum_{k=0}^{N-1} y_i[k]y_j^*[k+d\,(\mathrm{mod} \,N)]\\
&=C(d)\sum_{k=0}^{N-1} w_i[k]w_j^*[k+d]W_N^{-dk} \label{eq:pccf}
\end{align}
where $C[d]=W_N^{-(d+N\,(\mathrm{mod} \,2)+2q)d/2}$. We have exhaustively generated all QPP interleaved sequences for $N\le 128$ and Fig. 3 contains the magnitude of the PCCF at $d=0$ for the cases where $|\theta_{ij}(0)|<N$. This shows that for some $N$, it is possible to find QPPs such that the sequences $w_i[k]$ become orthogonal $(|\theta_{ij}(0)|/N=0)$. Notably, Fig. 3 also shows that there are cases, e.g., $N=98$, where no QPPs could be found to generate orthogonal sequences.

{\it Example 5:}
Consider $N=2^{n_0}$ with $n_0\geq 2$ and the two QPPs, $\pi_0[k]=2k^2+k\,(\mathrm{mod} \,N)$ and $\pi_1[k]=2k^2+\left ( \frac{N}{2}+1 \right ) k\,(\mathrm{mod} \,N)$. By inserting (\ref{eq:wmod}) in (\ref{eq:pccf}) with $d=0$ and with these QPPs, we obtain after some simplifications $|\theta_{01}[0]|=\sum_{k=0}^{N-1} (-1)^{k^2}=0$, i.e., $w_0[k]$ and $w_1[k]$ are mutually orthogonal. 

For a given sequence length $N$, the size $I$ of a set of orthogonal interleaved Zadoff-Chu sequences obtained by different QPPs was always such that $I<N$ for those values of $N \,(N\le 128)$ we have evaluated. For example, when $N=25$ we found $I=15$ and when $N=81$ we found $I=16$. Hence, we anticipate that a full set of orthogonal sequences (when $I=N$) cannot be found by merely using different QPPs. 

{\it Example 6:}
For $N=32$ it can be confirmed by evaluating (\ref{eq:pccf}) that a set of $I=4$ orthogonal sequences can be formed from the QPPs $\pi_0[k]=2k^2+k\,(\mathrm{mod} \,32)$,  $\pi_1[k]=2k^2+3k\,(\mathrm{mod} \,32)$, $\pi_2[k]=2k^2+17k\,(\mathrm{mod} \,32)$ and  $\pi_3[k]=2k^2+19k\,(\mathrm{mod} \,32)$. Letting $u=1$ and $q=0$, the modulation sequences (\ref{eq:mods}) have period of 16 elements and are equal to (\ref{eq:w0})-(\ref{eq:w3}) 
\begin{figure*}[tbp]
\hrulefill
{\small
\begin{align}
w_0[k\,(\mathrm{mod} \,16)]&=\left [1,\frac{1}{\sqrt{2}}-\frac{1}{\sqrt{2}}\mathrm{j},-1,\mathrm{j},1,\frac{1}{\sqrt{2}}+\frac{1}{\sqrt{2}}\mathrm{j},-1,-1,1,-\frac{1}{\sqrt{2}}+\frac{1}{\sqrt{2}}\mathrm{j},-1,-\mathrm{j},1,
-\frac{1}{\sqrt{2}}-\frac{1}{\sqrt{2}}\mathrm{j},-1,1\right]\label{eq:w0}\\
w_1[k\,(\mathrm{mod} \,16)]&=\left[1, -\frac{1}{\sqrt{2}}-\frac{1}{\sqrt{2}}\mathrm{j}, 1, -\mathrm{j}, 1, -\frac{1}{\sqrt{2}}+\frac{1}{\sqrt{2}}\mathrm{j},1,-1,1,\frac{1}{\sqrt{2}}+\frac{1}{\sqrt{2}}\mathrm{j},
1,\mathrm{j},1,\frac{1}{\sqrt{2}}-\frac{1}{\sqrt{2}}\mathrm{j}, 1,1\right]\label{eq:w1} \\
w_2[k\,(\mathrm{mod} \,16)]&=\left[1,-\frac{1}{\sqrt{2}}+\frac{1}{\sqrt{2}}\mathrm{j},-1,-\mathrm{j},1,-\frac{1}{\sqrt{2}}-\frac{1}{\sqrt{2}}\mathrm{j},-1,1,1,\frac{1}{\sqrt{2}}-\frac{1}{\sqrt{2}}\mathrm{j},-1,\mathrm{j},1,
\frac{1}{\sqrt{2}}+\frac{1}{\sqrt{2}}\mathrm{j},-1,-1\right]\label{eq:w2} \\
w_3[k\,(\mathrm{mod} \,16)]&=\left[1,\frac{1}{\sqrt{2}}+\frac{1}{\sqrt{2}}\mathrm{j},1,\mathrm{j},1,\frac{1}{\sqrt{2}}-\frac{1}{\sqrt{2}}\mathrm{j},1,1,1,-\frac{1}{\sqrt{2}}-\frac{1}{\sqrt{2}}\mathrm{j},1,-\mathrm{j},1,
 -\frac{1}{\sqrt{2}}+\frac{1}{\sqrt{2}}\mathrm{j}, 1, -1\right]\label{eq:w3} 
\end{align}
}%
\hrulefill 
\end{figure*}

\begin{figure}
\begin{centering}
\includegraphics[width=\linewidth]{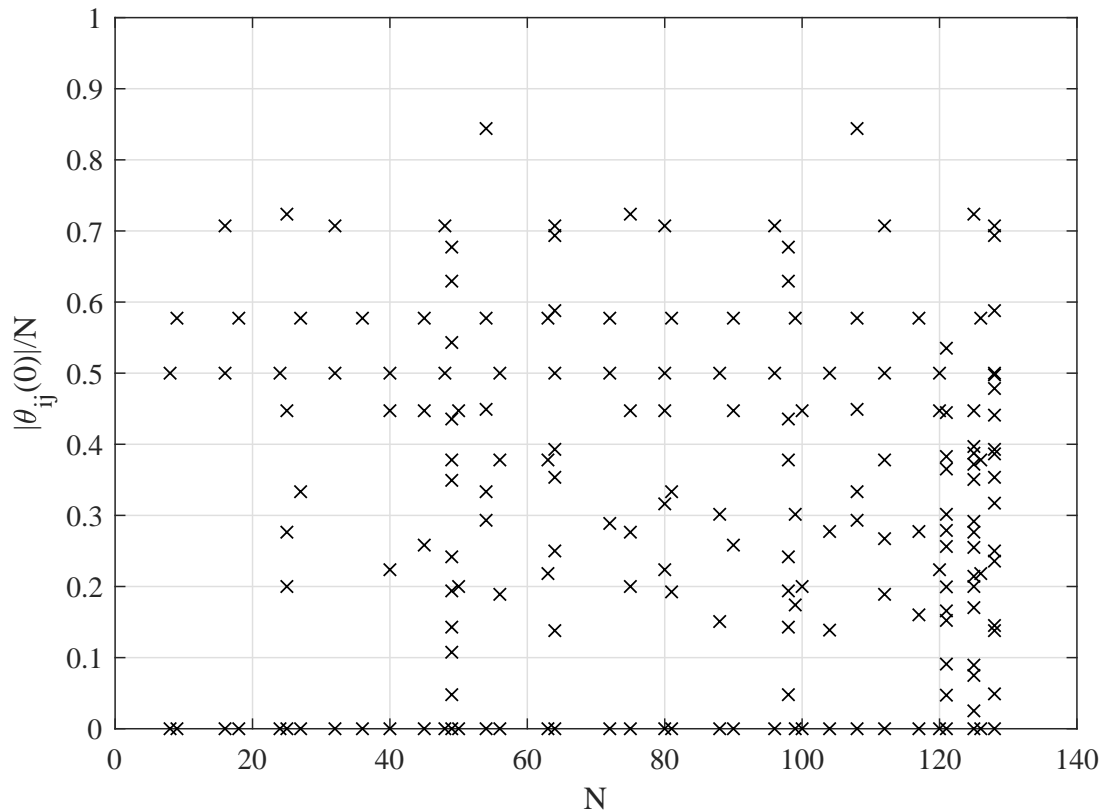}
\caption{Values of the magnitude of the normalized cross-correlation $|\theta_{ij}(d)|/N$ at $d=0$ for different $N$, for the interleaved Zadoff-Chu sequences constructed by using different QPPs and a common Zadoff-Chu sequence.}
\end{centering}
\label{fig:corr}
\end{figure}

\section{Conclusions}
A new family of CAZAC sequences is constructed from the permutation polynomial interleaved Zadoff-Chu sequences. This construction enriches the set of methods that can be used to generate CAZAC sequences. Interleaving of Zadoff-Chu sequences based on QPPs, or based on their inverse permutation polynomials, always preserves the CAZAC property. Permutation polynomials of order larger than two could be used as well, albeit the CAZAC property is not maintained generally. A set of orthogonal sequences can, in many cases, be obtained by using different QPPs. 
The interleaving of other families of CAZAC sequences might be an interesting topic for further research, as we observed that the Frank sequences  \cite{Frank} preserve the CAZAC property when interleaved by QPP permutations.

\appendices 
\section{Lemmas for Theorem 1.}
{\it Lemma 1:} If $\alpha$ is a positive integer, then 
\begin{align}
\mathrm{gcd}(\alpha f_2 d+f_1,N)=
\begin{cases}
2, & N\equiv 0 \, (\mathrm{mod}\, 2), N\nequiv 0 \, (\mathrm{mod}\, 4),\\
& f_2\equiv 1\, (\mathrm{mod}\, 2),d\equiv 0 \, (\mathrm{mod}\, 2)\\
1, & \mathrm{gcd}(p_i, f_2)\neq 1, 0\le i \le r. \label{eq:lm1}
\end{cases}
\end{align}
{\it Proof:} Suppose (\ref{eq:qppc1}) applies, then $N\equiv 0 \, (\mathrm{mod}\, 2)$, $N\nequiv 0 \, (\mathrm{mod}\, 4)$.  
From (\ref{eq:qppc1}) and the condition $f_2\equiv 1\, (\mathrm{mod}\, 2)$ it follows that $f_1\equiv0 \, (\mathrm{mod}\, 2)$, i.e.,  $\mathrm{gcd}(f_1,2)\neq 1$ while (\ref{eq:qppc3}) ensures that it is only 2 that divides both $f_1$ and $N$, so we have $\mathrm{gcd}(f_1, N)=2$. 
With the condition $d\equiv 0 \, (\mathrm{mod}\, 2)$ it follows that $\mathrm{gcd}(\alpha f_2 d+f_1,N)=2$.

Suppose (\ref{eq:qppc1}) applies with $f_2\equiv 0\, (\mathrm{mod}\, 2)$, or that  (\ref{eq:qppc2}) applies. Then (\ref{eq:qppc1})-(\ref{eq:qppc3}) give that $f_2\equiv 0\, (\mathrm{mod}\, p_i)$, i.e.,  $\mathrm{gcd}(p_i, f_2)\neq 1, 0\le i \le r$. Therefore, it further follows from (\ref{eq:qppc1})-(\ref{eq:qppc3}) that $f_1\nequiv 0\, (\mathrm{mod}\, p_i)$, i.e., $\mathrm{gcd}(p_i, f_1)= 1, 0\le i \le r$. Hence $p_i|f_2$, $p_i\nmid f_1$ and $p_i\nmid \alpha f_2 d+f_1$, i.e., $\mathrm{gcd}(\alpha f_2 d+f_1,N)=1$. $\hfill \blacksquare$

{\it Lemma 2:} If $\mathrm{gcd}(p_i, f_2)\neq 1, 0\le i \le r$ and 
\begin{equation}
t=N/\mathrm{gcd}(g_2,N)
\end{equation}
then,
\begin{align}
g_3t &\equiv 0 \, (\mathrm{mod}\, N) \label{eq:lemma3a}\\
g_2t&\equiv 0 \, (\mathrm{mod}\, N)\label{eq:lemma3b}\\
g_1t&\nequiv 0 \, (\mathrm{mod}\, N)\label{eq:lemma3c}
\end{align} 
where
\begin{align}
g_3&=2f_2^2d \\
g_2&=3f_2d(f_2d+f_1) \\
g_1&=d(2f_2d+f_1)(f_2d+f_1) +df_2(N\, (\mathrm{mod}\, 2)). 
\end{align}

{\it Proof:} To prove  (\ref{eq:lemma3a}) we have
\begin{align}
t&=\frac{N}{\mathrm{gcd}(3f_2d(f_2d+f_1),N)}\label{eq:L2a}\\
&=\frac{N}{\mathrm{gcd}(3f_2d,N)}\label{eq:L2b}
\end{align}
where the second equality follows from Lemma 1. Then we obtain
\begin{equation}
\mathrm{gcd}(3f_2d,N)=
\begin{cases}
\mathrm{gcd}(f_2d,N), & \mathrm{gcd}(3,N)=1 \label{eq:L2c}\\
3\mathrm{gcd}(f_2d,N/3), & \mathrm{gcd}(3,N)\neq 1. %\label{eq:L2d}
\end{cases}
\end{equation}
If $\mathrm{gcd}(3,N)\neq 1$, it follows from (\ref{eq:qppc3}) that $f_2\equiv 0 \, (\mathrm{mod}\, 3)$. Hence, from (\ref{eq:L2b}) and (\ref{eq:L2c}), we have
\begin{align}
g_3t&=
\begin{cases}
2f_2\frac{f_2d}{\mathrm{gcd}(f_2d,N)}N, & \mathrm{gcd}(3,N)=1\\
2\frac{f_2}{3}\frac{f_2d}{\mathrm{gcd}(f_2d,N/3)}N, & \mathrm{gcd}(3,N)\neq 1
\end{cases}
\\
&\equiv 0 \, (\mathrm{mod}\, N).
\end{align}

The proof of (\ref{eq:lemma3b}) follows directly from $g_2t=\frac{g_2N}{\mathrm{gcd}(g_2,N)} \equiv  0\, (\mathrm{mod}\, N)$. 

To prove (\ref{eq:lemma3c}), we can first write 
\begin{equation}
g_1t=\frac{g_1}{d}\frac{d}{\mathrm{gcd}(3f_2d,N)}N \label{eq:L2q}
\end{equation}
and we should show that $\frac{g_1}{d}\frac{d}{\mathrm{gcd}(3f_2d,N)}$ is not an integer.  Lemma 1 gives $\mathrm{gcd}((2f_2d+f_1)(f_2d+f_1),N)=1$ and therefore, $\mathrm{gcd}(g_1/d,N)=\mathrm{gcd}((2f_2d+f_1)(f_2d+f_1)+f_2,N)=1$. Thus, $\mathrm{gcd}(g_1/d,\mathrm{gcd}(3f_2d,N))=\mathrm{gcd}(\mathrm{gcd}(g_1/d,N),3f_2d)=1$, which implies it is sufficient to show that $\frac{d}{\mathrm{gcd}(3f_2d,N)}$ is not an integer. 

If $\mathrm{gcd}(d,N)=1$, then $\mathrm{gcd}(d,\mathrm{gcd}(3f_2d,N))=\mathrm{gcd}(\mathrm{gcd}(d,N),3f_2d)=1$ which implies that $d/\mathrm{gcd}(3f_2d,N)$ is not an integer, i.e., $g_1t\nequiv 0 \, (\mathrm{mod}\, N)$. 

If $\mathrm{gcd}(d,N)\neq 1$, suppose $N=p_0^{n_0} \ldots  p_r^{n_r}$ with $n_i>0$. Any valid $0< d \le N-1$ can be defined by 
$d=d'p_0^{\alpha_0} \ldots  p_r^{\alpha_r}$ with $\alpha_i\ge 0$ and  $\mathrm{gcd}(d',N)=1$.  Moreover, any $f_2$ can be defined by $f_2=f'p_0^{\beta_0} \ldots  p_r^{\beta_r}$  with $\mathrm{gcd}(f',N)=1$, where $\beta_i>0$ due to (\ref{eq:qppc3}).  
Then $\mathrm{gcd}(3f_2d,N)=p_0^{\gamma_0} \ldots  p_r^{\gamma_r}$ with $\gamma_i=\min (\alpha_i+\beta_i,n_i)$ if $p_i\neq 3$ and $\gamma_i=\min (\alpha_i+\beta_i+1,n_i)$ if $p_i=3$.
To prove $g_1t\nequiv 0 \, (\mathrm{mod}\, N)$, it should be shown that 
$\frac{d}{\mathrm{gcd}(3f_2d,N)}=d'p_0^{\alpha_0-\gamma_0} \ldots  p_r^{\alpha_r-\gamma_r}$ is not an integer.
If $\gamma_i=\alpha_i+\beta_i$ and $p_i\neq 3$, or $\gamma_i=\alpha_i+\beta_i+1$ and $p_i=3$, then $\alpha_i<\gamma_i$ since $\beta_i>0$. 
If $\gamma_i=n_i$ and if there is a $j$ for which  $\alpha_j\ge \gamma_j$, then there exists at least one $i \,(i\neq j)$ for which $\alpha_i<\gamma_i$.  
This follows from $d<N$, which implies that there exists at least one $i$ for which $\alpha_i<n_i$, and thus $\gamma_i>\alpha_i$. 
Hence, $\frac{d}{\mathrm{gcd}(3f_2d,N)}$ is not an integer and thus $g_1t\nequiv 0 \, (\mathrm{mod}\, N)$.$\hfill \blacksquare$

{\it Lemma 3:} If $N\equiv 0 \, (\mathrm{mod}\, 2)$, $N\nequiv 0 \, (\mathrm{mod}\, 4)$, $f_2 \equiv 1\, (\mathrm{mod}\, 2)$ and
\begin{equation}
t=
\begin{cases}
N/\mathrm{gcd}(g_2,N), & d\equiv 0 \, (\mathrm{mod}\, 2)\\
N/\mathrm{gcd}(2g_2,N), & d\equiv 1 \, (\mathrm{mod}\, 2)
\end{cases}
\end{equation}
then,
\begin{align}
g_3t &\equiv 0 \, (\mathrm{mod}\, N) \label{eq:lemma4a}\\
2g_2t & \equiv 0 \, (\mathrm{mod}\, N) \label{eq:lemma4c}\\
g_2t^2+g_1t&\nequiv 0 \, (\mathrm{mod}\, N)\label{eq:lemma4b}
\end{align}
where
\begin{align}
g_3&=2f_2^2d \label{eq:L3g3}\\
g_2&=3f_2d(f_2d+f_1) \label{eq:L3g2}\\
g_1&=d(2f_2d+f_1)(f_2d+f_1) +df_2(N\, (\mathrm{mod}\, 2)). \label{eq:L3g1}
\end{align} 

{\it Proof:} To prove (\ref{eq:lemma4a}) for $d\equiv 1 \, (\mathrm{mod}\, 2)$, we have 
\begin{align}
t&=\frac{N}{\mathrm{gcd}(6f_2d(f_2d+f_1),N)}\\
&=\frac{N}{2\mathrm{gcd}(3f_2d,N/2)} \label{eq:L3a}
\end{align}
where (\ref{eq:L3a}) follows directly from $\mathrm{gcd}(f_2d+f_1,N)=1$, which holds because
$\mathrm{gcd}(f_2d,2)=1$ and $\mathrm{gcd}(f_1,N)=2$ according to (\ref{eq:qppc1}) and (\ref{eq:qppc3}). 
If $\mathrm{gcd}(3,N)\neq 1$, it follows from (\ref{eq:qppc3}) that $f_2\equiv 0 \, (\mathrm{mod}\, 3)$ and we obtain
\begin{align}
g_3t&=
\begin{cases}
f_2\frac{f_2d}{\mathrm{gcd}(f_2d,N/2)}N, & \mathrm{gcd}(N,3)=1\\
\frac{f_2}{3}\frac{f_2d}{\mathrm{gcd}(f_2d,N/6)}N, & \mathrm{gcd}(N,3)\neq 1
\end{cases}
\\
&\equiv 0 \, (\mathrm{mod}\, N).
\end{align}
For $d\equiv 0 \, (\mathrm{mod}\, 2)$, 
\begin{align}
t&=\frac{N}{\mathrm{gcd}(3f_2d(f_2d+f_1),N)}\\
&=\frac{N}{\mathrm{gcd}(6f_2d,N)} \label{eq:L3tval}
\end{align}
where (\ref{eq:L3tval}) comes from $\mathrm{gcd}(f_2d+f_1,N)=2$ according to Lemma 1.
Hence, we obtain
\begin{align}
g_3t&=
\begin{cases}
f_2\frac{2f_2d}{\mathrm{gcd}(2f_2d,N)}N, & \mathrm{gcd}(N,3)=1\\
\frac{f_2}{3}\frac{2f_2d}{\mathrm{gcd}(2f_2d,N/3)}N, & \mathrm{gcd}(N,3)\neq 1
\end{cases}
\\
&\equiv 0 \, (\mathrm{mod}\, N).
\end{align}

To prove (\ref{eq:lemma4c}), we directly obtain 
\begin{align}
2g_2t&=
\begin{cases}
\frac{2g_2N}{\mathrm{gcd}(2g_2,N)}, & d\equiv 1 \, (\mathrm{mod}\, 2)\\
2\frac{g_2N}{\mathrm{gcd}(g_2,N)}, & d\equiv 0 \, (\mathrm{mod}\, 2)
\end{cases}
\\
&\equiv 0 \, (\mathrm{mod}\, N).
\end{align}

To prove (\ref{eq:lemma4b}) for $d\equiv 1 \, (\mathrm{mod}\, 2)$, it suffices to show that 
\begin{equation}
g_2t^2+g_1t \nequiv 0 \, (\mathrm{mod}\, 2) \label{eq:L3c1}
\end{equation}
since $N\equiv 0 \, (\mathrm{mod}\, 2)$ and thus $g_2t^2+g_1t$ must be divisible by 2, if it is divisible by $N$.

To prove (\ref{eq:L3c1}), we start from $g_2=3f_2d(f_2d+f_1)\equiv 1 \, (\mathrm{mod}\, 2)$, since $f_2\equiv 1 \, (\mathrm{mod}\, 2)$ and $d\equiv 1 \, (\mathrm{mod}\, 2)$. We also have $t=\frac{N}{\mathrm{gcd}(2g_2,N)}=\frac{N/2}{\mathrm{gcd}(g_2,N/2)}\equiv 1 \, (\mathrm{mod}\, 2)$ since $N/2$ is odd. Thus $g_2t^2\nequiv 0 \, (\mathrm{mod}\, 2)$. As $f_2\equiv 1 \, (\mathrm{mod}\, 2)$, from (\ref{eq:qppc1}) it follows $f_1\equiv 0 \, (\mathrm{mod}\, 2)$, so from (\ref{eq:L3g1}) we obtain $g_1 \equiv 0 \, (\mathrm{mod}\, 2)$. Thus, it follows $g_2t^2+g_1t \nequiv 0 \, (\mathrm{mod}\, 2)$.
 
For $d\equiv 0 \, (\mathrm{mod}\, 2)$, we have $g_2t=\frac{g_2N}{\mathrm{gcd}(g_2,N)}\equiv 0 \, (\mathrm{mod}\, N)$ and therefore $g_2t^2\equiv 0 \, (\mathrm{mod}\, N)$. Thus, it remains to show that $g_1t\nequiv 0 \, (\mathrm{mod}\, N)$. Lemma 1 gives $(2f_2d+f_1)(f_2d+f_1)=4\epsilon$ where $\mathrm{gcd}(\epsilon,N)=1$. Thus from (\ref{eq:L3g1}) and (\ref{eq:L3tval}) we obtain 
\begin{align}
g_1t&=\epsilon\frac{4d}{\mathrm{gcd}(6f_2d,N)}N. \label{eq:L3g1t}
\end{align}
It should be shown that $\frac{4d}{\mathrm{gcd}(6f_2d,N)}$ is not an integer. 
According to the assumptions of Lemma 3, $N=2p_1^{n_0} \ldots  p_r^{n_r}$ with $n_i>0$ and $p_i\neq 2$ for $i=1,2,\ldots,r$. Any even $d$ where $0< d \le N-1$ can be defined by 
$d=d'2^{\alpha_0}p_1^{\alpha_1} \ldots  p_r^{\alpha_r}$ with $\alpha_i\ge 0$ and $\alpha_0>0$, and  $\mathrm{gcd}(d',N)=1$.
According to the condition of Lemma 3, we consider any odd $f_2$ which can be defined by $f_2=f'p_1^{\beta_1} \ldots  p_r^{\beta_r}$ with $p_i\neq2$, $\mathrm{gcd}(f',N)=1$ where $\beta_i>0$ due to (\ref{eq:qppc3}).  
Then (\ref{eq:L3g1t}) becomes
\begin{align}
g_1t&=\epsilon d'\frac{2^{\alpha_0+1}p_1^{\alpha_1}\ldots p_r^{\alpha_r}}{\mathrm{gcd}(3p_1^{\beta_1}\ldots p_r^{\beta_r}p_1^{\alpha_1}\ldots p_r^{\alpha_r}, p_1^{n_0} \ldots  p_r^{n_r})}N\\
&=\epsilon d' 2^{\alpha_0+1}p_1^{\alpha_1-\gamma_1}\ldots p_r^{\alpha_r-\gamma_r}N
\end{align}
with $\gamma_i=\min (\alpha_i+\beta_i,n_i)$ if $p_i\neq 3$ and $\gamma_i=\min (\alpha_i+\beta_i+1,n_i)$ if $p_i=3$. 
If $\gamma_i=\alpha_i+\beta_i$ and $p_i\neq 3$, or $\gamma_i=\alpha_i+\beta_i+1$ and $p_i=3$, then $\alpha_i<\gamma_i$ since $\beta_i>0$. 
If $\gamma_i=n_i$ and if there is a $j$ for which  $\alpha_j\ge \gamma_j$, then there exists at least one $i \,(i\neq j)$ for which $\alpha_i<\gamma_i$.  
This follows from $d<N$ and $d\equiv 0 \, (\mathrm{mod}\, 2)$, which implies that there exists at least one $i$ for which $\alpha_i<n_i$, and thus $\gamma_i>\alpha_i$. 
Hence, $2^{\alpha_0+1}p_1^{\alpha_1-\gamma_1}\ldots p_r^{\alpha_r-\gamma_r}$ is not an integer and $g_1t\nequiv 0 \, (\mathrm{mod}\, N)$.$\hfill \blacksquare$

\section{Lemma for Theorem 2.}

\emph{Lemma 4:} If $\pi^{-1}[k]=h_2k^2+h_1k+h_0\, (\mathrm{mod} N)$ is a permutation polynomial and $\mathrm{gcd}(u,N)=1$, then $u+2kh_2 \nequiv 0 \, (\mathrm{mod} N)$ for any integer $k$.

\emph{Proof:} It  follows from  the conditions on the QPP coefficients (\ref{eq:qppc1})-(\ref{eq:qppc3}) that there exists at least one $p_i$ for which
\begin{equation}
h_2 \equiv 0 \, (\mathrm{mod}\, p_i),  \label{eq:g2}
\end{equation} 
which implies that 
\begin{equation}
\mathrm{gcd}(h_2,N)\neq 1.
\end{equation}
Assume that $u+2kh_2 \equiv 0 \, (\mathrm{mod} N)$, which implies that there exists an integer $s$ such that 
\begin{equation}
u+2kh_2 =sN
\end{equation} 
which means that $(sN-2kh_2)/\mathrm{gcd}(h_2,N)$ is an integer, which implies that $\mathrm{gcd}(u,\mathrm{gcd}(h_2,N))\neq 1$. However, we have
\begin{align}
\mathrm{gcd}(u,\mathrm{gcd}(h_2,N))&=\mathrm{gcd}(\mathrm{gcd}(u,N),h_2)\\
&= \mathrm{gcd}(1,h_2)\\
&=1
\end{align}
which is a contradiction. Hence, $u+2kh_2 \neq sN$.$\hfill \blacksquare$

%\begin{IEEEbiographynophoto}
%{Fredrik Berggen} (Senior Member, IEEE)
%received the M.Sc. degree in applied physics and
%electrical engineering from Linköping University,
%Sweden, in 1998, and the Ph.D. degree in electrical
%engineering from the Royal Institute of
%Technology, Sweden, in 2003. 
%Since 2005, he has been with Huawei Technologies, Stockholm,
%Sweden, where he holds the position as Expert.
%He participated in 3GPP RAN1 for more than
%a decade and has contributed to physical layer
%standardization of 4G Long Term Evolution (LTE) and 5G New Radio (NR). He is the inventor of numerous
%standard-essential patents. His research interests include radio access
%networks, with emphasis on physical layer waveforms, channels, and procedures.
%Dr. Berggren is a co-recipient of the IEEE Communications Society
%Heinrich Hertz Award for Best Communications Letter in 2010.
%\end{IEEEbiographynophoto}
%\begin{IEEEbiographynophoto}
%{Branislav M. Popovi\'{c}} 
%received the Ph.D.
%degree in electrical engineering from the School
%of Electrical Engineering, University of Belgrade,
%Serbia. Prior to joining Huawei Technologies,
%Stockholm, Sweden, in 2001, he was with
%the Institute of Microwave Techniques and
%Electronics, Belgrade, from 1984 to 1994,
%Ericsson, Stockholm, from 1994 to 2000, and
%Marconi, Stockholm, from 2000 to 2001. He is
%a Huawei Fellow.
%\end{IEEEbiographynophoto}
\end{document}